\documentclass[conference]{IEEEtran}
\IEEEoverridecommandlockouts
\usepackage{cite}
\usepackage{amsmath,amssymb,amsfonts}
\usepackage{algorithmic}
\usepackage{graphicx}
\usepackage{xcolor}
\usepackage{import}
\usepackage{color}
\usepackage{caption}
\usepackage{subcaption}
\usepackage{textcomp}
\usepackage{array}
\usepackage{pgfplots}
\usepackage{listings}
\usepackage{comment}
\def\BibTeX{{\rm B\kern-.05em{\sc i\kern-.025em b}\kern-.08em
    T\kern-.1667em\lower.7ex\hbox{E}\kern-.125emX}}

\graphicspath{{fig/}}
\definecolor{light-gray}{gray}{0.80}
\definecolor{dark-red}{RGB}{150, 54, 52}
\definecolor{dark-blue}{RGB}{54, 96, 146}

\pgfplotsset{compat=1.11,
    /pgfplots/ybar legend/.style={
    /pgfplots/legend image code/.code={%
       \draw[##1,/tikz/.cd,yshift=-0.25em]
        (0cm,0cm) rectangle (3pt,0.8em);},
   },
}

\begin{document}

\title{System-on-Chip Security Assertions\\
}

\author{\IEEEauthorblockN{Yangdi Lyu and Prabhat Mishra}
\IEEEauthorblockA{\textit{Department of Computer and Information Science and Engineering} \\
\textit{University of Florida, Gainesville, Florida, USA}
}
}

\maketitle

\begin{abstract}
Assertions are widely used for functional validation as well as coverage analysis for both software and hardware designs. Assertions enable runtime error detection as well as faster localization of errors. While there is a vast literature on both software and hardware assertions for monitoring functional scenarios, there is limited effort in utilizing assertions to monitor System-on-Chip (SoC) security vulnerabilities. In this paper, we identify common SoC security vulnerabilities by analyzing the design. To monitor these vulnerabilities, we define several classes of assertions to enable runtime checking of security vulnerabilities. Our experimental results demonstrate that the security assertions generated by our proposed approach can detect all the inserted vulnerabilities while the functional assertions generated by state-of-the-art assertion generation techniques fail to detect most of them.
\end{abstract}


\section{Introduction}
System-on-Chip (SoC) security is critical as more and more personal computing needs as well as physical infrastructures are controlled by a chip with the prevalence of Internet-of-Things (IoTs). Attackers take advantage of security vulnerabilities to inject malicious software. Tools such as anti-virus are not enough to protect from these attacks.
Security vulnerabilities can arise in any stage, from design to fabrication, as well as post deployment. 
In contrast to a software vulnerability which can be modified after deployment, fixing a hardware vulnerability becomes more and more difficult and expensive in later stages. Existing approaches try to mask some of the vulnerabilities using firmware patching or utilizing in-built reconfigurable primitives. However, these approaches may not work in all scenarios. Therefore, detecting and removing vulnerabilities in early stages is very important in SoC designs.

Assertion-Based Verification (ABV) is a common practice in industry today for functional validation of System-on-Chip (SoC) designs \cite{foster2004assertion}. Assertions define the properties that should hold. For example, a functional assertion can check that the output of an adder is equal to the sum of two inputs irrespective of the implementation. In addition to checking the inputs and outputs, assertions can also increase the observability of internal states, which enables runtime error detection as well as faster localization of errors. 
While  there  is  a  vast  literature  on both software and hardware assertions for monitoring functional scenarios, there is limited effort in utilizing assertions to monitor SoC security vulnerabilities~\cite{Ray:2018}.

\begin{figure}
    \centering
    \includegraphics[width=\linewidth]{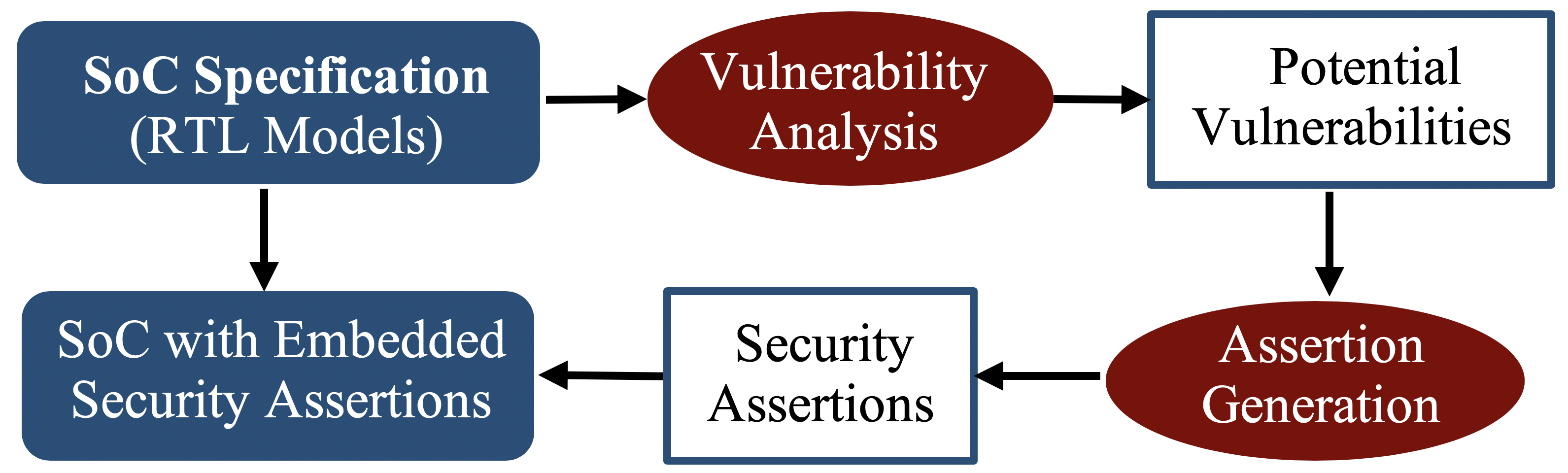}
    \caption{Our proposed framework consists of two main steps. First, we perform vulnerability analysis on a given SoC to identify potential vulnerabilities. Next, assertions are generated based on the vulnerabilities and inserted into the design.}
    \label{fig:overview}
    \vspace{-0.2in}
\end{figure}

Existing SoC validation techniques mainly focus on the functional behaviors defined in the specification. Traditionally, SoC security vulnerabilities are not considered as expected functional behaviors. For example, an unspecified transition in finite state machine (FSM) is one of the main sources of security vulnerabilities. While ABV is the de facto standard for functional validation, there are limited prior efforts to define and monitor SoC security vulnerabilities. Given the importance of SoC security, we propose a framework for defining and utilizing SoC security assertions as shown in Figure~\ref{fig:overview}. The framework consists of two main steps. First, we perform vulnerability analysis on a given SoC design and identify potential vulnerabilities. Next, security assertions are generated from the vulnerabilities and inserted into the SoC design. The purpose of this paper is not to provide a laundry list of vulnerabilities for SoC designs. Instead, we identify some representative vulnerabilities and propose assertions for them to show how SoC security assertions can be defined and integrated in an existing SoC validation methodology. 
Specifically, this paper makes three major contributions:
\begin{enumerate}
    \item We perform a comprehensive review of the literature to identify the common SoC security vulnerabilities.
    \item We propose security assertions corresponding to each vulnerability. We plan to upload the security assertions for various open source SoCs in TrustHub \cite{TrustHub}.
    \item We demonstrate that existing assertion-based validation is not capable of detecting security vulnerabilities. Therefore, security assertions defined in this paper should be part of any security validation methodology.
\end{enumerate}

The remainder of this paper is organized as follows. Section~\ref{sec:background} surveys related approaches. Section~\ref{sec:vulner} provides an overview of SoC security vulnerabilities. Section~\ref{sec:assert} describes the framework for assertion generation for a given set of security vulnerabilities. Section~\ref{sec:case} presents six case studies. Finally, Section~\ref{sec:conclusion} concludes the paper.

\section{Background and Related Work}
\label{sec:background}
There are significant prior efforts in classifying software-level vulnerabilities \cite{cwe,nvd}. While there are related works in specific areas (e.g., classifying hardware Trojans), to the best of our knowledge, there are limited previous works on developing a comprehensive classification of potential SoC vulnerabilities.
Similarly, there are various efforts on functional assertions in the context of assertion-based validation. However, there are limited prior efforts in defining and utilizing SoC security assertions. In this section, we present the related works in assertion-based validation as well as automated generation of assertions.

\subsection{Assertion-based Validation}
There are mainly two types of approaches for defining hardware assertions: language-based and library-based \cite{di2013integration}. Language-based approaches provides syntax for formally defining assertions. Two of the most popular assertion specification languages are Property Specification Language (PSL) \cite{ieee2010psl} and System Verilog Assertions (SVA) \cite{ieee2012sva}. Both of these languages support temporal assertions and formally is an extension of temporal logic \cite{ben1983temporal}. Some other examples are ForSpec \cite{armoni2002forspec}, SALT \cite{bauer2011theory}, a SystemC extension \cite{tabakov2008temporal}, etc. On the other hand, library-based approaches adds assertion support to existing languages. One such example is Open Verification Library (OVL) \cite{foster2006introduction}. OVL has support for Verilog, VHDL, PSL and SystemVerilog. Library-based approaches can be used to quickly write common types of assertions. Unfortunately, they are not generic enough to cover all possible scenarios. In this paper, we use SVA to express our SoC security assertions.

\subsection{Automated Assertion Generation}
Assertion generation is mostly manual effort - it is time-consuming to insert enough assertions into an industrial SoC design. Many research efforts have been devoted to automated generation of functional assertions. Rogin et al. \cite{Rogin:2008} proposed to generate properties of a design by analyzing simulation traces. 
Hertz et al. \cite{Hertz:2013} improved the analysis process using data mining and developed a tool named Goldmine. The generated rules from simulation traces are passed through a formal verification tool to verify the correctness in the design. As the simulation data is inherently incomplete and nondeterministic, the quality of mined assertions cannot be guaranteed. Moreover, as our case studies show, these functional assertions are not suitable for detecting  security vulnerabilities.


\section{SoC Security Vulnerabilities}
\label{sec:vulner}
While there is a vast literature on both software and hardware assertions for monitoring functional scenarios, there are limited efforts to define and utilize assertions to monitor System-on-Chip (SoC) security vulnerabilities~\cite{Ray:2018, Wang:2018}. We reviewed a wide variety of security vulnerabilities listed in the National Vulnerability Database \cite{nvd} and related research~\cite{Ray:2018}, and identified common SoC security vulnerabilities that are related to SoCs. Note that there is a fundamental difference between exceptions and security vulnerabilities. The exceptions are defined today by SoC designers based on the point of view of functional correctness, whereas the security vulnerabilities outlined in this proposal are solely from security and trust perspectives. For example, an adversary may trigger (e.g., using a Trojan) an exception (e.g., divide by zero) solely to gain a higher privilege level, such that private memory or registers can be accessed. The remainder of this section describes our proposed seven vulnerability classes.   

\subsection{Permissions and Privileges} 
Permissions and privileges are the main components of the access control subsystem. Specifically, different resources are controlled by different permissions and privileges. For example, in ARM7 processor, seven different modes are defined, such as user mode, interrupt mode, and supervisor mode. It is critical to check whether the conditions for triggering privileged modes are satisfied before
changing modes.

\subsection{Resource Management}
Certain resources should be protected against any illegal
access, including accessing special hardware from non-privileged
modes, misuse of design-for-debug infrastructures
during normal usage, and so on. For example, JTAG allows
engineers to trace secure memory during post-silicon validation and debug of security features.
However, JTAG should never be enabled during normal usage.

\subsection{Illegal States and Transitions}
The behavior of an SoC can be modeled as a finite state machine (FSM). The valid states or transitions can be verified during functional validation. Attackers are more interested in the backdoor that allows undefined states/transitions. To verify the existence of illegal states and transitions, we could use assertions of valid states and transitions to alarm any violation, or enumerate all possible invalid states and transitions to prevent specific vulnerabilities. 

\subsection{Buffer Issues}
Modern SoCs consist of advanced features (e.g., out-of-order execution and speculative execution) as well as a large number of heterogeneous buffers. Similar to software buffer errors,
these buffers in deeper pipelines require significant validation efforts to detect any remaining flaws.
For example, prefetched instructions in buffers should be flushed if the branch prediction is incorrect. Otherwise, these flaws can be exploited to mount an attack.

\subsection{Information Leakage}
Modern SoCs try to isolate a secure world from a non-secure
world. Information from the secure world should never
be leaked to non-secure world directly. ARM uses TrustZone
as an approach to provide the secure world \cite{trustzone}. There should be safeguards 
present to prevent non-secure world from accessing TrustZone
directly.

\subsection{Numeric Exceptions}
Numeric exceptions represent the erroneous/illegal behaviors (e.g., divide by zero) during arithmetic computations. Even if the program does not lead to illegal numeric computation, an attacker can make it happen, and utilize it to create a vulnerability. 

\subsection{Malicious Implants}
In software community, code injection means allowing
attackers to run arbitrary code. Similarly, hardware Trojans, inserted by untrusted third party, allow attackers to
execute an arbitrary path after applying specific input patterns. This can lead to information leakage or other unintended consequences.
Trojans are usually inserted in hard-to-detect and rare-to-activate areas, making it hard to detect them during validation.

\vspace{-0.2in}
\begin{figure}[h]
    \centering
    \includegraphics[width=\linewidth]{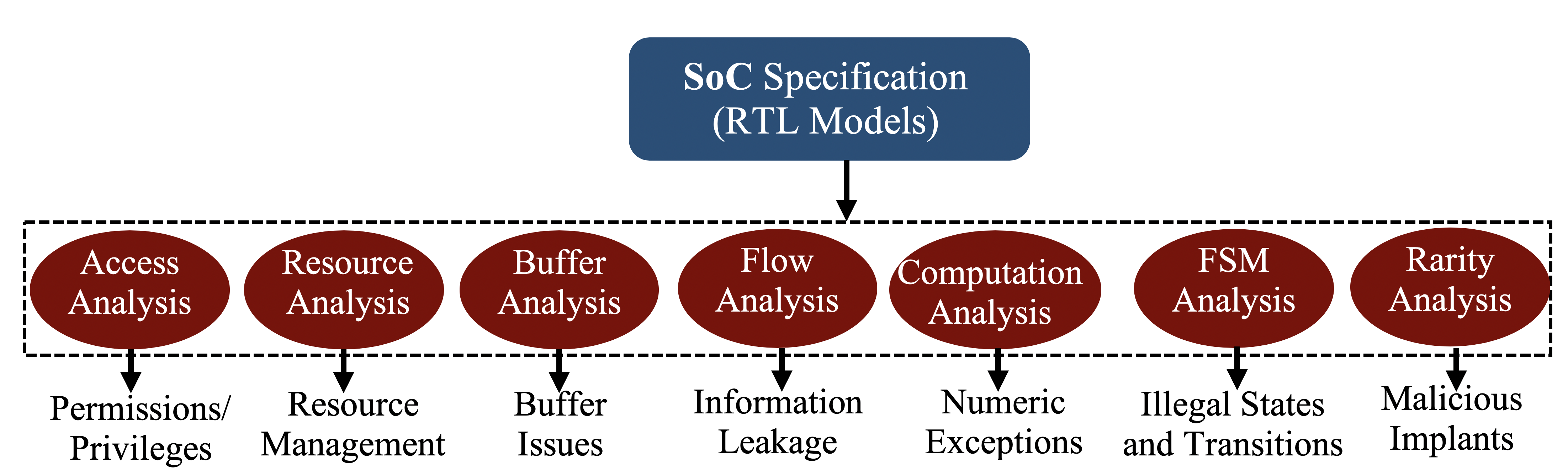}
    \caption{Overview of our proposed assertion generation framework for the seven classes of SoC vulnerabilities.}
    \label{fig:classes}
    \vspace{-0.1in}
\end{figure}
\section{SoC Security Assertions}
\label{sec:assert}
This section mainly addresses two important questions: i) how to generate the security assertions, and ii) how to embed them in the SoC design (RTL models).

\subsection{Embedding of Security Assertions} 
There are two orthogonal ways of embedding assertions in the RTL model of an SoC design: immediate and concurrent assertions. 
Please note that immediate assertions can be converted to concurrent assertions by modifying the antecedent. However, as described below, it would be natural to use a specific one depending on the type of security vulnerability. 

\vspace{0.05in}
\noindent {\it Immediate Assertions:} Immediate assertions are powerful in detecting vulnerabilities such as numeric errors. Immediate assertions are flexible and can vary based on the potential statements or blocks. For immediate operations, it is important to find out the exact location, the relevant variable (e.g., trigger) $\alpha$, and the assertion, {\it assert ($P(\alpha)$)}, can be inserted. Immediate assertions are inherent for checking specific operations, such as divide-by-zero checking and out-of-boundary checking.

\vspace{0.05in}
\noindent {\it Concurrent Assertions:} Concurrent assertions, on the other hand, are checked each clock cycle, representing expected properties of SoCs. Concurrent assertions are useful to express any FSM related vulnerabilities (e.g., illegal states and transitions). Each concurrent assertion can be defined using {\it assert property} $(P)$. The property $P$ should be derived from the specification of SoCs and vulnerability classes.

\subsection{Generation of Security Assertions}
We use static (code) analysis to generate security assertions to detect the existence of the vulnerabilities outlined in Section~\ref{sec:vulner}, as shown in Figure~\ref{fig:classes}. In this section, we briefly outline the assertion generation for each vulnerability class.  

\vspace{0.05in}
\noindent {\it Permissions and Privileges:}
By analyzing the specification, we need to figure out the variable that represents the privilege level, e.g., CPSR in ARMv7. For each entry to a privileged operation block, we need to generate an assertion. For the ease of illustration, we use \texttt{user} to represent current privilege, and \texttt{admin} to represent root privilege. For each possible entry into the privileged operation block, we need to generate the immediate assertion as: $assert (\texttt{user == admin})$.

\vspace{0.05in}
\noindent {\it Resource Management:}
Concurrent assertions are powerful in protecting resources from misuse in an unexpected way. For example, to protect JTAG from getting used during normal operation mode, we need to generate the assertion as: {\it assert property} \texttt{(normal $|->$ !JTAG\_enable)}. 

\vspace{0.05in}
\noindent {\it Illegal States and Transitions:}
The behaviors of modules as well as their interactions (protocols) can be expressed in FSMs. Therefore, we can express both valid and invalid (illegal) transitions as security assertions. For example, if there is a valid 
transition from state $A$ to state $B$ when variable $a$ is true, it can be
encoded as an assertion: {\it assert property (A \&\& a $|->$  B)}. 
Similarly, the assertion, {\it assert property} (\texttt{!}$(C |-> A)$)
can be used to ensure that no transitions are allowed from state $C$ 
to state $A$.


\vspace{0.05in}
\noindent {\it Buffer Issues: }
Assertions can be generated for all boundary cases related to buffers. To prevent access of the buffer index beyond its limits, immediate assertions should be added before each buffer access. Before accessing \texttt{Buffer[index]}, the variable \texttt{index} needs to satisfy {\it assert} \texttt{(index >= 0 \&\& index <= limit)}.
In many scenarios, we may require concurrent assertions to ensure global interactions. For example, to ensure the flush of instruction buffer (IB) after branch prediction failure, we can use the assertion {\it assert property} \texttt{(Pre\_fail $|->$ IB\_Empty)}.

\vspace{0.05in}
\noindent {\it Information Leakage:}
To protect secret information from directly leaking to non-secure world, tagging is one potential solution. It assumes that the results of secure world can only be passed to non-secure world through special interface (privileged instructions). For each normal operation consisting of both secure and non-secure variables, we need to check if the result is assigned to a non-secure variable as expected. If $s$ is a secure variable, the assignment of $s$ to a variable $v$ needs to check the tag of $v$ as {\it assert(v\_t == secure\_tag)}.

\vspace{0.05in}
\noindent {\it Numeric Exceptions:}
Numeric exceptions are more relevant to the implementation of SoC designs. We need to generate one assertion for each possible numeric exception during arithmetic computation. For example, in case of a divide-by-zero exception, we can generate an immediate assertion as: {\it assert}\texttt{(divisor != 0)}.

\vspace{0.05in}
\noindent {\it Malicious Implants:}
Malicious modifications (e.g., hardware Trojans) can be inserted during pre-silicon or post-silicon stage. In the pre-silicon stage, hardware Trojans are usually hidden in rare-to-activate branches or rare execution of concurrent statements. For example, we can generate assertions for each rare branch by adding an assertion for each rare trigger condition as: {\it assert(rare\_trigger)}. 

The security assertions are generated based on the  vulnerabilities outlined in Section~\ref{sec:vulner}. More assertions can be added based on designer inputs or application-specific considerations.

\vspace{-0.2in}
\begin{figure}[h]
\center
{
\footnotesize
\pgfplotstableread[row sep=\\,col sep=&]{
    benchmarks & Our & Goldmine \\
    Arbiter & 100 & 100\\
    PCI & 100 & 20\\
    USB & 100 & 10\\
    MEM & 100 & 0\\
    GNG & 100 & 0\\
    AES & 100 & 0\\
    }\mydata

\begin{tikzpicture}
\footnotesize
    \begin{axis}[
            ybar,
            bar width=.3cm,
            width=0.95\linewidth,
            height=.6\linewidth,
            legend style={at={(0.5,1)},
                anchor=north,legend columns=-1, draw=none, 
                /tikz/column 2/.style={
                column sep=15pt,
                },
            },
            symbolic x coords={Arbiter,PCI,USB,MEM,GNG,AES},
            xtick=data,
            ymin=0,ymax=120,
            ytick={0, 20, 40, 60, 80, 100},
            yticklabel={\pgfmathprintnumber{\tick}\%},
            ylabel={\% of Detected Vulnerabilities},
        ]
        \addplot[fill=dark-red] table[x=benchmarks,y=Our]{\mydata};
        \addplot[fill=dark-blue] table[x=benchmarks,y=Goldmine]{\mydata};
        \legend{Our approach, Goldmine}
    \end{axis}
\end{tikzpicture}
}
\caption{Comparison of our assertions and assertions generated by Goldmine \cite{Hertz:2013} in detecting security vulnerabilities.}
\label{fig:compare}
\vspace{-0.2in}
\end{figure}
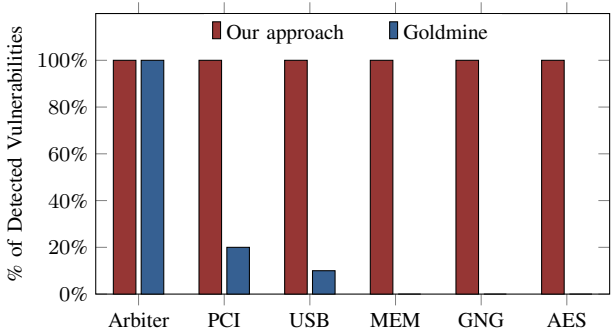

\section{Case Study}
\label{sec:case}

To demonstrate the necessity of security assertions, we analyzed six benchmarks and inserted security assertions introduced in Section~\ref{sec:assert}. Then, Goldmine \cite{Hertz:2013} was applied on all the benchmarks to generate as many assertions as possible. To evaluate the effectiveness of our security assertions, we randomly inserted 10 vulnerabilities into the design to form 10 vulnerable instances and applied directed test to activate these vulnerabilities. If any assertion generated by the two methods (ours versus Goldmine) got activated during simulation, we claim the corresponding method detects the vulnerability. The types of potential vulnerabilities of each benchmark and the detected instances are shown in Table~\ref{tab:result} and Figure~\ref{fig:compare}, respectively. Note that the number of instances are more than the number of vulnerability types, as each type may contain multiple instances. In the remainder of this section, we describe each type of vulnerability and inserted instances in detail. Overall, our approach is able to detect all of the vulnerabilities while the assertions generated by Goldmine fail to detect most of them.

\begin{table}[h]
    \centering
    \begin{tabular}{|p{2cm}|c|c|c|c|c|c|}
        \hline
        Vulnerability & Arbiter & PCI & USB & MEM & GNG & AES\\
        \hline
        Permissions and Privileges & & & & \checkmark  & & \\
        \hline
        Resource Management & & & \checkmark & & & \checkmark \\
        \hline
        Illegal States \& Transitions & \checkmark & \checkmark & \checkmark & & &\\
        \hline
        Buffer Issues & & & & \checkmark & &\\
        \hline
        Information Leakage & & & & \checkmark  & &\\
        \hline
        Numeric Exceptions & & & & & \checkmark &\\
        \hline
        Malicious Implants & & & & & & \checkmark\\
        \hline
    \end{tabular}
    \caption{Types of vulnerabilities explored in the six benchmarks (more potential vulnerabilities are possible).}
    \label{tab:result}
    \vspace{-0.2in}
\end{table}


\subsection{Arbiter}
\begin{figure}
    \centering
    \includegraphics[width=.8\linewidth]{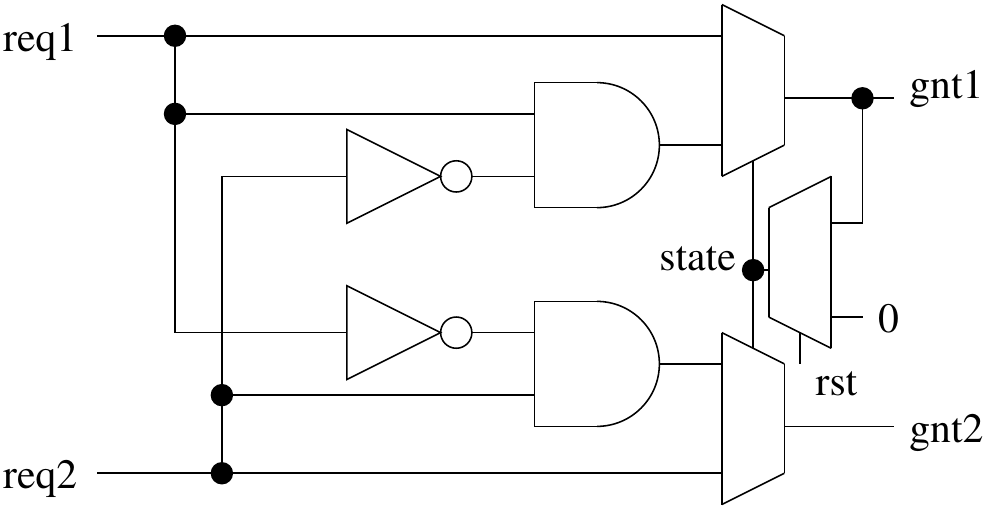}
    \caption{A simple arbiter with four inputs (req1, req2, rst, clk (not shown)) and two outputs (gnt1, gnt2).}
    \label{fig:arbiter}
    \vspace{-0.2in}
\end{figure}

We first analyzed a simple design, Arbiter, as shown in Figure~\ref{fig:arbiter}. For this simple design, we inserted the vulnerability of {\bf invalid states and transitions}. Even for this small design, careless design, fault injections, or transient errors can make it behave differently. For example, if the security of whole design relies on the {\it gnt1} and {\it gnt2} not asserted together, {\it assert}(!(gnt1 \& gnt2)) should be added to the design. However, the number of invalid states and transitions would be exponential when time is involved. For example, when we consider two consecutive cycles, {\it assert}(!(gnt1 $|=>$ (gnt2 != req2))) should hold. We limit the number of assertions to 10 in this example.
  





The assertions generated by Goldmine are shown in Listing~\ref{lst:assertions}. As Goldmine analyzes the golden design by mining the traces of random simulation, it is able to capture reachable states and transitions by the specific simulation.

\vspace{-0.1in}
\begin{center}
\begin{minipage}{3.3in}
\begin{lstlisting}[caption={Assertions of Arbiter by Goldmine}, label={lst:assertions}]
(state == 1 & req2 == 1) |-> (gnt1 == 0)
(req1 == 1 & state == 0) |-> (gnt1 == 1)
(req1 == 0) |-> (gnt1 == 0)
(req1 == 1 & req2 == 0) |-> (gnt1 == 1)
(req1 == 1 & state == 0) |-> (gnt2 == 0)
(req2 == 1 & state == 1) |-> (gnt2 == 1)
(req2 == 0) |-> (gnt2 == 0)
(req2 == 1 & req1 == 0) |-> (gnt2 == 1)
\end{lstlisting}
\end{minipage}
\end{center}
The 10 vulnerability instances are generated by mimicking the behavior of fault injection, i.e., randomly inverting one signal in Figure~\ref{fig:arbiter}. As we can see from Figure~\ref{fig:compare}, Goldmine is good at detecting vulnerabilities involving finite state machines in this small design. 

\subsection{PCI}
Top module pci\_master32\_sm from Opencores \cite{OpenCores} contains eight modules such as pci\_frame\_crit and pci\_irdy\_out\_crit. To mimic SoC design, which contains different parts from untrusted third party, we inserted {\bf invalid states and transitions} to the subordinate modules (8 out of 10) as well as the top modules (2 out of 10). Goldmine generated 19 assertions. To generate vulnerable instances, we randomly changed operators in all the modules. As shown in Figure~\ref{fig:compare}, Goldmine was able to capture the two vulnerabilities, but failed to detect the remaining eight vulnerabilities.

\subsection{USB Protocol}
USB protocol defines the packet fields and its corresponding operations. We analyzed the USB protocol module usbf\_pa.v along with CRC module from Opencores \cite{OpenCores} and identified two types of vulnerabilities. The USB protocol depends on the packet ID (PID) to identify the types of packets including token, data, handshake, and special. For each type of packets, PIDs will not overlap. In the module, the output of PID is stored in tx\_data whose value depends on the input selectors. The two types of vulnerabilities are shown below:
\begin{enumerate}
    \item {\bf Resource management:} As the PIDs define the resources of USB, it is critical to make sure that each type of packet gets expected PID. For example, a token packet should not get any of the PIDs belonging to data packets, such as DATA0, DATA1. Similarly, a handshake packet should stick to its own type, e.g., ACK, NAK. We inserted 8 assertions of this type.
    \item {\bf Invalid states and transitions: } We inserted one vulnerability in CRC module and one vulnerability in the top module. As this is a common vulnerability in almost every design, we will skip inserting this type of vulnerability in the remaining benchmarks.
\end{enumerate}
The vulnerable instances were generated accordingly. As shown in Figure~\ref{fig:compare}, while Goldmine can only detect one vulnerability from state transition in top module, our assertions can detect all of them.

\subsection{A Simplified Memory Design}
This design is created to mimic the behavior of a simplified Trusted Hardware (TH) implementation of memory, as shown in Listing~\ref{lst:memory}. Trusted Hardware, e.g., Intel’s Software Guard Extensions (SGX) \cite{sgx}, allows remote clients to upload private computation and data to a secure container of a server with a TH. One key implementation of SGX is the introduction of Process Reserved Memory and Enclave Page Cache (EPC), inhibiting invalid accesses even from the kernel. The simplified design is shown in Listing~\ref{lst:memory} which contains input signal {\it sc} to denote whether it is a secure access or not. The memory space is denoted by an array named {\it mem} with size of 1MB.

\begin{enumerate}
    \item {\bf Permissions and privileges}. Assume the lower 1kB of memory is allocated to EPC. Since EPC should be accessed through secure container/process, each access to the lower 1kB should be checked. Although one conditional checking is already in place, assertions may also help when implementation error, fault injection or Hardware Trojans exist, e.g., {\it assert}(address $<=$ 1024 $|=>$ sc) can be inserted before any access to memory.
    \item {\bf Information leakage}. In this simplified memory implementation, it does not explicitly describe the state of {\it out} signal when we want to write. For a buggy CPU design which connects to this memory, a process may be able to read the previous access of another process from the out port (including secure processes) with interleaved memory access. We may add a concurrent assertion with ({\it assert property} (wr $|=>$ out == 0)).
    \item {\bf Buffer errors}. Memory as one type of buffer should be checked for buffer errors. Each access to memory should be checked with assertions to test if address is in the range of memory size. In this example, the memory size is 1MB ($2^{20}$) and the length of address is 32 bits. We need {\it assert}(address $< 2**20$) to check out-of-boundary accesses.
\end{enumerate}

\vspace{-0.1in}
\begin{center}
\begin{minipage}{3.3in}
\begin{lstlisting}[frame=single, caption={Simplified Memory of Trusted Hardware}, basicstyle=\small, language=Verilog, label={lst:memory}]
module mem(clk, rst, wr, sc, address,
            in, out);
input clk, rst, wr, sc;
input [31:0]address;
input [7:0]in;
output reg [7:0]out;
reg[7:0] mem[2**20-1:0];
always @ (posedge clk)
  if (address >= 1024 || sc) begin
    if (wr) mem[address] <= in;
    else out <= mem[address];
  end
endmodule
\end{lstlisting}
\end{minipage}
\end{center}

We inserted a vulnerability related to permissions and privileges, by removing {\it sc} checking in the first {\it if} statement. For vulnerabilities related to information leakage, we assume that attackers are able to connect one specific location to {\it out} when it is a write operation. Experimental results show that Goldmine failed to detect any of these vulnerabilities while our approach can detect all of them.

\vspace{-0.1in}
\begin{center}
\begin{minipage}{3.3in}
\begin{lstlisting}[frame=single, caption={GNG\_interp}, language=Verilog, basicstyle=\small, label={lst:gng}]
module gng_interp (
  input clk, rstn, valid_in,
  input [63:0] data_in,
  output reg valid_out, 
  output reg [15:0] data_out
);
wire [33:0] mul1;
wire signed [13:0] mul1_new;
reg [17:0] c0_r5;
reg signed [18:0] sum2;
reg [14:0] sum2_rnd;
assign mul1_new = mul1[32:19];
always @ (posedge clk)
  sum2 <= $signed({1'b0, c0_r5}) + mul1_new;
always @ (posedge clk)
  sum2_rnd <= sum2[17:3] + sum2[2];
...
endmodule
\end{lstlisting}
\end{minipage}
\end{center}

\subsection{Gaussian Noise Generator (GNG)}
We next inspected a computation-intensive design called Gaussian Noise Generator (GNG). The design is downloaded from Opencores \cite{OpenCores}. One possible {\bf numeric error} in this design is the assignments between signed and unsigned values as shown in the snippet of code in Listing~\ref{lst:gng}. The first assignment assigns an unsigned variable to a signed variable. Next assignment is the computation between signed values. The final assignment assigns a signed value to an unsigned value. We are concerned with the automatic transformation between signed and unsigned values. For example, when the $32^{th}$ bit of mul1 is 1, mul1\_new is interpreted as a negative value using a two's complement representation. Then mul\_new is added to a positive number and converted to an unsigned number again. The behavior may or may not be the original intention of this code. We want to generate assertions, e.g., {\it assert}(mul1[32] != 1), and guide the test plan of debug. The developer should decide if it is a numeric error or the expected behavior.  
Since we view this design as ``buggy" by itself, we did not generate vulnerable instances for it. Rather, we use 10 direct tests to force mul1[32] to become 1 and check if the assertions by Goldmine can catch the potential vulnerability. As shown in Figure~\ref{fig:compare}, Goldmine failed to detect any of them since it only analyzes the traces of random simulation but never inspects the specification/implementation. 

\vspace{-0.1in}
\begin{center}
\begin{minipage}{3.3in}
\begin{lstlisting}[frame=single, caption={AES table}, language=Verilog, basicstyle=\small, label={lst:aes}]
module S4 (clk, JTAG, in, out, JTAG_out);
  input clk, JTAG;
  input [31:0] in;
  output [31:0] out;
  output reg [31:0] JTAG_out;
  S
    S_0 (clk, in[31:24], out[31:24]),
    S_1 (clk, in[23:16], out[23:16]),
    S_2 (clk, in[15:8],  out[15:8] ),
    S_3 (clk, in[7:0],   out[7:0]  );
  always @ (posedge clk)
    if (JTAG) JTAG_out <= in;
endmodule
module S (clk, in, out);
  always @ (posedge clk)
  case (in)
    8'h00: out <= 8'h63;
    8'h01: out <= 8'h7c;
    ...
  endcase
endmodule
\end{lstlisting}
\end{minipage}
\end{center}

\subsection{AES}
Advanced Encryption Standard (AES) is a very commonly used crypto core,  consisting of ten rounds of block ciphers (substitution permutation networks). The substitution is shown in Listing~\ref{lst:aes}. We also inserted JTAG to dump internal variables during debug. The identified vulnerabilities are:
\begin{enumerate}
    \item {\bf Resource management: } As JTAG is for debug purpose only, it should be disabled during normal usage. As a result, the dump signal should contains nothing related to any internal signals. We inserted an concurrent assertion {\it assert property} \texttt{(!JTAG |-> (JTAG\_out == 0))} to prevent attackers from bypassing the JTAG checking and dump internal signals to infer the plaintext.
    \item {\bf Malicious implants: } As module S contains a lot of rare branches, attackers are able to construct rare trigger conditions for hardware Trojans. For example, the probability of (out == 32'h7c7c7c7c) is $2^{-32}$ when (in == 8'h01) is true for all S\_0, S\_1, S\_2 and S\_3 together. Assertions like {\it assert}(out != 32'h7c7c7c7c) in module S4 can guide the designer of a test plan to cover this specific potential trigger condition. As the combinations of rare branches are potentially infinite, we restricted the number of assertions to be 10.
\end{enumerate}
One of our vulnerable instances is a design bypassing JTAG checking directly. For the other 9 instances, we construct random hardware Trojans from the rare branches. As shown in Figure~\ref{fig:compare}, our approach is able to detect the vulnerabilities. Goldmine cannot detect any of them. 

Overall, the security assertions generated by our approach are able to detect all the security vulnerabilities whereas the assertions generated by state-of-the-art technique (Goldmine) fail to detect most of them.

\section{Conclusion}
\label{sec:conclusion}
SoCs are widely used today in both embedded systems and IoT devices. While SoC security is paramount, there are limited prior efforts in defining and detecting a wide variety of SoC security vulnerabilities. In this paper, we developed seven classes of SoC security vulnerabilities. Based on these vulnerabilities, we proposed a framework for generating security assertions. Using a diverse set of benchmarks, we demonstrated that the functional assertions generated by state-of-the-art assertion generation technique cannot eliminate the need for our dedicated security assertions. Specifically, our security assertions are able to detect all the implanted security vulnerabilities while the  state-of-the-art method failed to detect most of them. We envision that the SoC designers will embed security assertions in their designs in the near future as part of their assertion-based security validation methodology. This will open up several research directions in terms of how to generate automated tests to activate these security assertions as well as how to utilize these security assertions (properties) for automated property checking. 
\bibliographystyle{IEEEtran}
\bibliography{references.bib}

\begin{thebibliography}{10}
\providecommand{\url}[1]{#1}
\csname url@samestyle\endcsname
\providecommand{\newblock}{\relax}
\providecommand{\bibinfo}[2]{#2}
\providecommand{\BIBentrySTDinterwordspacing}{\spaceskip=0pt\relax}
\providecommand{\BIBentryALTinterwordstretchfactor}{4}
\providecommand{\BIBentryALTinterwordspacing}{\spaceskip=\fontdimen2\font plus
\BIBentryALTinterwordstretchfactor\fontdimen3\font minus
  \fontdimen4\font\relax}
\providecommand{\BIBforeignlanguage}[2]{{%
\expandafter\ifx\csname l@#1\endcsname\relax
\typeout{** WARNING: IEEEtran.bst: No hyphenation pattern has been}%
\typeout{** loaded for the language `#1'. Using the pattern for}%
\typeout{** the default language instead.}%
\else
\language=\csname l@#1\endcsname
\fi
#2}}
\providecommand{\BIBdecl}{\relax}
\BIBdecl

\bibitem{foster2004assertion}
H.~D. Foster, A.~C. Krolnik, and D.~J. Lacey, \emph{Assertion-based
  design}.\hskip 1em plus 0.5em minus 0.4em\relax Springer Science \& Business
  Media, 2004.

\bibitem{Ray:2018}
S.~{Ray}, E.~{Peeters}, M.~M. {Tehranipoor}, and S.~{Bhunia}, ``System-on-chip
  platform security assurance: Architecture and validation,'' \emph{Proceedings
  of the IEEE}, vol. 106, no.~1, pp. 21--37, Jan 2018.

\bibitem{TrustHub}
``Trusthub,'' \url{https://www.trust-hub.org/}, accessed: 2018-10-10.

\bibitem{cwe}
``Common weakness enumeration,'' \url{https://cwe.mitre.org/}, accessed:
  2018-10-10.

\bibitem{nvd}
``National vulnerability database,'' https://nvd.nist.gov.

\bibitem{di2013integration}
G.~Di~Guglielmo, L.~Di~Guglielmo, A.~Foltinek, M.~Fujita, F.~Fummi,
  C.~Marconcini, and G.~Pravadelli, ``On the integration of model-driven design
  and dynamic assertion-based verification for embedded software,''
  \emph{Journal of Systems and Software}, vol.~86, no.~8, 2013.

\bibitem{ieee2010psl}
``{1850-2010 - IEEE Standard for Property Specification Language (PSL)},''
  2010.

\bibitem{ieee2012sva}
``{1800-2012 - IEEE Standard for SystemVerilog--Unified Hardware Design,
  Specification, and Verification Language},'' 2012.

\bibitem{ben1983temporal}
M.~Ben-Ari, A.~Pnueli, and Z.~Manna, ``The temporal logic of branching time,''
  \emph{Acta informatica}, vol.~20, no.~3, pp. 207--226, 1983.

\bibitem{armoni2002forspec}
R.~Armoni, L.~Fix, A.~Flaisher, R.~Gerth, B.~Ginsburg, T.~Kanza, A.~Landver,
  S.~Mador-Haim, E.~Singerman, A.~Tiemeyer \emph{et~al.}, ``The forspec
  temporal logic: A new temporal property-specification language,'' in
  \emph{International Conference on Tools and Algorithms for the Construction
  and Analysis of Systems}.\hskip 1em plus 0.5em minus 0.4em\relax Springer,
  2002, pp. 296--311.

\bibitem{bauer2011theory}
A.~Bauer and M.~Leucker, ``The theory and practice of salt,'' in \emph{NASA
  Formal Methods Symposium}.\hskip 1em plus 0.5em minus 0.4em\relax Springer,
  2011, pp. 13--40.

\bibitem{tabakov2008temporal}
D.~Tabakov, G.~Kamhi, M.~Y. Vardi, and E.~Singerman, ``A temporal language for
  systemc,'' in \emph{Formal Methods in Computer-Aided Design, 2008.
  FMCAD'08}.\hskip 1em plus 0.5em minus 0.4em\relax IEEE, 2008, pp. 1--9.

\bibitem{foster2006introduction}
H.~Foster, K.~Larsen, and M.~Turpin, ``Introduction to the new accellera open
  verification library,'' in \emph{DVCon’06: Proceedings of the Design and
  Verification Conference and exhibition}.\hskip 1em plus 0.5em minus
  0.4em\relax Citeseer, 2006.

\bibitem{Rogin:2008}
F.~Rogin, T.~Klotz, G.~Fey, R.~Drechsler, and S.~Rulke, ``Automatic generation
  of complex properties for hardware designs,'' in \emph{2008 Design,
  Automation and Test in Europe}, March 2008, pp. 545--548.

\bibitem{Hertz:2013}
S.~Hertz, D.~Sheridan, and S.~Vasudevan, ``Mining hardware assertions with
  guidance from static analysis,'' \emph{IEEE Transactions on Computer-Aided
  Design of Integrated Circuits and Systems}, vol.~32, no.~6, pp. 952--965,
  June 2013.

\bibitem{Wang:2018}
C.~{Wang}, Y.~{Cai}, Q.~{Zhou}, and H.~{Wang}, ``Asax: Automatic security
  assertion extraction for detecting hardware trojans,'' in \emph{2018 23rd
  Asia and South Pacific Design Automation Conference (ASP-DAC)}, Jan 2018, pp.
  84--89.

\bibitem{trustzone}
``Arm trustzone,'' https://developer.arm.com/technologies/trustzone.

\bibitem{OpenCores}
``Opencores,'' \url{https://www.opencores.org/}, accessed: 2018-11-10.

\bibitem{sgx}
``Intel® software guard extensions,'' https://software.intel.com/en-us/sgx.

\end{thebibliography}

\end{document}